# Reliability assessment in advanced nanocomposite materials for orthopaedic applications

Jérôme Chevalier[a], Paola Taddei[b], Laurent Gremillard[a], Sylvain Deville[a], Gilbert Fantozzi[a], J. F. Bartolomé[c], C. Pecharroman[c], J.S. Moya[c], L.A. Diaz[d], R. Torrecillas[d], Saverio Affatato[e]

[a] Université de Lyon, INSA Lyon, MATEIS - UMR CNRS 5510,
69621 Villeurbanne Cedex, France
[b] Centro di Studio sulla Spettroscopia Raman, Dipartimento di Biochimica "G. Moruzzi", Sezione di Chimica e Propedeutica Biochimica,
Via Belmeloro 8/2, Università di Bologna, 40126 Bologna, Italy
[c] ICMM, Spanish Research Council (CSIC), Campus Canto Blanco, Madrid, Spain
[d] INCAR, Spanish Research Council (CSIC),
La Corredoria s/n Ap. 73, 33080 Oviedo, Spain
[e] Laboratorio di Tecnologia Medica, Istituto Ortopedico Rizzoli, Bologna, Italy

## Abstract

Alumina-zirconia nano-composites were recently developed as alternative bearing materials for orthopedics. Previous, preliminary reports show that such alumina-zirconia nanocomposites exhibit high crack resistance and low wear rate. In this paper, additional information is given in terms of wear, crack resistance and ageing behaviour: femoral heads are inspected after 7 million cycles of wear testing on a hip simulator, crack resistance is measured and compared to other ceramics used today in orthopedics, slow crack growth is reported under static and cyclic fatigue, and aging resistance is assessed. We also report on the load to failure of femoral heads prototypes during compression tests. This overall reliability assessment ensures a potential future development for these kinds of new nanocomposites in the orthopedic field.

# 1. Introduction

Nowadays, the main issue for THR is the generation of wear debris produced mainly by the acetabular component (Campbell et al., 2004). Standard artificial hip joints consist of a polymer cup made of Ultrahigh Molecular Weight Polyethylene (UHMWPE), placed in the acetabulum via a metal-back component, and a metal (stainless steel or cobalt chromium alloy) ball fixed to a metal stem introduced in the femur. Any use of the joint results in cyclic stress of the polymer cup against the metal ball. During the reciprocating motion of normal joint use, UHMWPE fibrils are then released, mostly by adhesive wear, to form billions of pieces of sub-micrometer wear debris that shed into the surrounding synovial fluid and tissues with negative biological effects. It is currently believed that the UHMWPE particles generated at the contact surfaces enter the periprosthetic tissues where they trigger a macrophage reaction. Then macrophages release pro-inflammatory cytokines that stimulate osteoclastic bone resorption, leading to osteolysis and eventual loosening of the prosthesis and a need for revision. This has led to the development of alternative surfaces for hip articulation such as ceramic-on-UHMWPE, metal-on-metal and ceramic-on-ceramic couplings. In the former coupling, the amount of ion released can be detrimental to the surrounding tissues; in comparison, the use of bio-ceramic materials reduces wear rate and produces no metal ion release. Due to the low roughness exhibited by ceramic heads (typically lower than a few nanometres) and their good wettability, abrasion is reduced significantly if ceramic femoral heads are used with acetabular cups made of polyethylene and almost completely avoided when using ceramic femoral heads together with ceramic cup inserts.

The clinical success associated to the use of ceramics led to the implantation of more than 3.5 million alumina components and more than 600.000 zirconia femoral heads worldwide since 1990, with a strongly growing market (i.e. more than 25% growth for the alumina–alumina coupling between 2002 and 2004). There are many reports on fracture rates associated with ceramics. Important references have been compiled elsewhere (Campbell et al., 2004). If, in the pioneering days, the fracture rate was quite high (up to 13% for some series), the *in vivo* failure rate reported by the producer of Biolox® alumina is today below 0.01% (Willmann, 2000). A comparable failure rate was claimed by the producer of Prozyr® zirconia heads (Cales, 2000). Critical events in 2001 are discussed below.

If the clinical follow up with current alumina ceramics is significant, it must be kept in mind that their use has been restricted so far to a limited number of designs of hip components, for which the mechanical loading is less demanding. As an example, 22 mm heads are not currently manufactured using alumina ceramics, for reliability reasons. Alumina–alumina coupling also exhibits a significant increase of failure rates. This is again related to its modest mechanical properties. In the 90's, Yttria-Stabilized Zirconia became a popular alternative to alumina as structural ceramic because of substantially higher fracture toughness and strength. The use of zirconia has opened the way towards new implant designs that were not possible with alumina which was too brittle. Biomedical grade zirconia exhibits the best mechanical properties among single-phase oxide ceramics: this is the consequence of phase conversion toughening, which increases its crack propagation resistance. The stress-induced phase transformation involves the transformation of metastable tetragonal grains to the monoclinic phase at the crack tip. It is accompanied by volume expansion and induces compressive stresses that hinder crack propagation. On the other hand, due to this meta-stability, zirconia is prone to aging in the presence of water (Chevalier, 2006). Zirconia manufacturers considered this problem as a minor issue until 2001, when roughly 400 failures of zirconia heads were reported within a very short period. The failure origin is now associated to an accelerated aging in two particular batches. Even if limited in time and number, and clearly identified to be process controlled, these events have had a catastrophic impact for the use of zirconia, some surgeons returning to other solutions. More important, some clinical reports show that zirconia can exhibit a progressive ageing degradation even under 'normal' situations, which limits the long-term stability of zirconia (Chevalier et al., 2007).

For these different reasons, there is a trend to develop stronger and more reliable ceramics that should expand the field of application of bio-inert ceramics in orthopaedics. Alumina–zirconia micro-composites (with both alumina and zirconia grains in the micrometer range) are today under development in the orthopaedic community and show improvement in mechanical properties and aging resistance (De Aza et al., 2002; Deville et al., 2003; Willmann, 1998). In a previous work, new alumina–zirconia composites, with an alumina matrix in the micrometer range and nano-sized zirconia particles have been developed (Chevalier et al., 2005). In the best cases (1.7 vol.% of intra-granular zirconia nano-particles in alumina), they exhibited fracture resistance properties that were never reached with oxide ceramics before. Wear tests were also run on a hip simulator at 7 million cycles and show that wear rates of these composites were comparable to alumina–alumina couplings, ensuring excellent wear properties as compared to metal–metal or metal–polyethylene coupling (Affatato et al., 2006). Following these promising preliminary results, our aim was therefore to give a complete picture of the wear resistance and mechanical properties of these nanocomposites, associated to long-term durability assessment.

Our aim was therefore first to inspect the surface of the components already tested at 7 million cycles on a hip simulator to investigate possible damage after wear.

If wear and potential debris release is one key issue for hip joint replacement, the potential of a new ceramic material must also include its sensitivity to slow crack growth and delayed failure, which is not taken into account in the sole strength and toughness data. As a general trend, the susceptibility of ceramics to Slow Crack Growth (SCG) is discussed on the basis of a $V$(crack velocity) versus $K_I$ (stress intensity factor) diagram ($K_I$ representing the stresses at the tip of a crack or any pre-existing defect such as a pore, a scratch, etc., in the ceramic). Recently, the presence of a threshold in the stress intensity factor, under which no crack propagation occurs, has been the subject of important research in the ceramic field (Wan et al., 1990). This threshold corresponds to equilibrium with null crack velocity. For ceramic joint prostheses for example, this threshold, $K_{I0}$, determines a safety range of use (De Aza et al., 2002). Due to the tendency of ceramics to undergo cyclic fatigue degradation (i.e. a decrease of $K_{I0}$ and an increase of crack rates), Slow Crack Growth (SCG) analysis must include cyclic fatigue experiments, which is rarely done in the bio-ceramics literature (Attaoui et al., 2005; Chevalier et al., 1999c; Dauskardt et al., 1994). Our aim was therefore to characterize deeply the SCG resistance under static and cyclic fatigue of the present nanocomposite material in comparison to biomedical grade alumina, zirconia and to a conventional alumina–zirconia micro-composite.

The addition of alumina to zirconia at least reduces drastically aging kinetics. It is shown that the strength of Biolox delta® for example does not decrease even when repeatedly steam sterilized. However, 'no decrease in strength' does not necessarily mean 'no aging', since other manifestations of ageing are grain pull out and roughening (Chevalier, 2006; Chevalier et al., 2007). Few studies have been devoted to aging in alumina–zirconia systems, but they show that, even if limited and possibly reduced to zero, some degree of degradation can be observed, depending on microstructural features. As an example, we showed in a previous work (Pecharroman et al., 2003) that aging could be significant in a 3Y-TZP–alumina composite above 16 vol.% zirconia. This critical content was related to the percolation threshold above which a continuous path of zirconia grains allowed transformation to proceed. The presence of aggregates in the microstructure may be also detrimental (Gutknecht et al., 2007). Any extrapolation to other laboratory scale or industrial composites could be hazardous, but it shows how aging must be checked carefully prior to clinical development of a given alumina–zirconia composite.

Accelerated aging tests were therefore conducted on the present material to ensure its perfect aging resistance.

At last, the development of a new material for hip joint applications must include the measurement of mechanical properties on real components. Therefore, we report here on the load to failure of femoral head prototypes during compression tests for a specific design where commercial alumina fail to follow ISO specifications.

## 2. Materials and methods

### 2.1. Materials preparation

The processing route to obtain the nano-composites specimens was described in detail elsewhere (Chevalier et al., 2000; Schehl et al., 2002). It consists in doping a stable suspension of a high purity alumina powder (Condea HPA 0.5, with an average particle size of 0.45 µm and a surface area of 10 m$^2$/g) in ethanol absolute (99.97%) by drop wise addition of a diluted (2/3 vol.% Zr alkoxide, 1/3 vol.% ethanol absolute) zirconium alkoxide (Aldrich Zirconium-IV-propoxide 70 wt% solution in 1-propanol). In the present work, a low amount of zirconia precursor was added, in order to obtain composites with only 1.7 vol.% (2.5 wt%) zirconia nano-particles. After drying under magnetic stirring at 70°C, the powders were thermally treated at 850°C for 2 h to remove organic residues and were subsequently attrition milled with alumina balls for 1 h. Green compacts were then obtained by a pressure-casting method. The optimum sintering to obtain the desired nano-structural distribution of zirconia particles consisted of a thermal treatment of 1600 °C/2 h. Fig. 1 shows the microstructure of the material, consisting in zirconia nano-particles ($D_{50}$≈150nm) evenly distributed in the alumina matrix ($D50$≈5µm). Those zirconia particles were found to be mainly (>70%) intragranular, with almost perfect spherical shape. These particles are well below the critical size for phase transformation (Schehl et al., 2002). All ceramics were fully dense.

### 2.2. Inspection of worn ceramic components after 7 million cycles

Twelve 28 mm×44 mm ceramic femoral heads, processed under the conditions mentioned above, were articulated with twelve ceramic acetabular cups of the same composition. The following configurations were tested on a hip simulator under bovine calf serum as a lubricant:

- four commercially available alumina heads and acetabular cups (Biolox forte® in the following referred to as AL);
- four experimental pure alumina heads and acetabular cups (in the following referred to as BK);
- four experimental nanocomposite heads and acetabular cups (in the following referred to as NK).

Major details about this experimental in-vitro test are available in literature (Affatato et al., 2006).

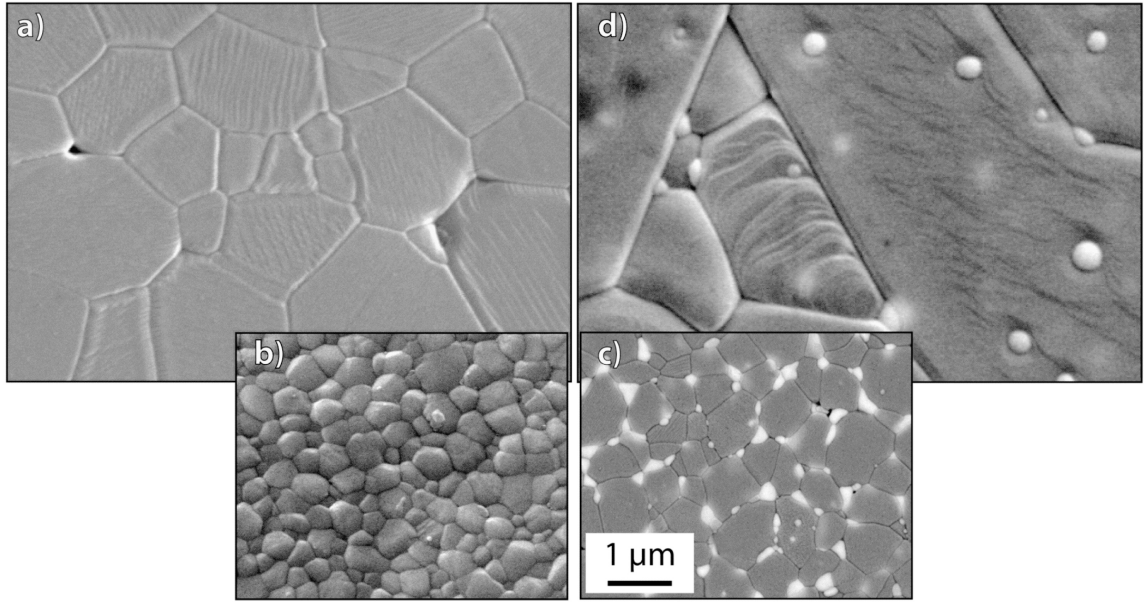

Fig. 1. Microstructure of monolithic alumina (a), yttria stabilized zirconia (b), conventional alumina–zirconia composite (c) and of the advanced nano-composite material (d).

## 2.3. Scanning Electron and Atomic Force Microscopy observations

Scanning Electron Microscopy (SEM, XL-20, FEI, Netherlands) observations were conducted at the surface of a commercial alumina head (AL), an experimental alumina head (BK) and a NK head after 7 million cycles. The heads were coated with a gold layer of 10 nm prior to observations.

Atomic Force Microscopy (AFM, D3100, Digital Instruments) measurements were conducted at the surface of the same heads. The Atomic Force Microscope (AFM) was used in contact mode, with an oxide-sharpened silicon nitride probe exhibiting a tip radius of curvature of 20 nm. The scans were performed at the surface with an average scanning speed of 10 µm/s. The scans were all conducted on the same day (i.e. with the same probe, with the same state) to lower the scatter of the data and to obtain comparative results. Different zones of the same head were observed in order to get statistical data. A computer-assisted treatment was conducted to suppress the roundness of the heads and to obtain Ra (roughness) values. These values were compared to those measured on the as-processed heads, i.e. before the wear tests.

## 2.4. Photoluminescence measurements

The surface residual stress state in the acetabular cups was measured via a piezospectroscopic technique, widely applied to the study of alumina and alumina–zirconia composites (Garcia et al., 2002; Ma and Clarke, 1994, 1993; Merlani et al., 2001; Sergo et al., 1998). The piezospectroscopic effect may be defined as the shift, induced by strain, in the frequency of a spectral band. The bands here considered to evaluate the residual stress state were the $R_1$ and $R_2$ fluorescence bands (at about 14 400 cm$^{-1}$ and 14 430 cm$^{-1}$, respectively), which have a long known and well-defined stress dependence (Garcia et al., 2002; Ma and Clarke, 1994, 1993; Merlani et al., 2001; Sergo et al., 1998). Photoluminescence is due to the radiative electronic transitions of the $Cr^{3+}$ ions, naturally present in alumina ceramics as trace impurities which substitute the $Al^{3+}$ ions in the $Al_2O_3$ lattice (He and Clarke, 1995; Selcuk and Atkinson, 2002). The origin of the piezospectroscopic effect is that when the lattice of ions surrounding the $Cr^{3+}$ is distorted, for instance by an applied stress, the crystal field potential at the site of the $Cr^{3+}$ ion is altered, which, in turn, alters the energies of the electronic transitions. Thus, the analysis of the fluorescence spectrum of $Cr^{3+}$-doped alumina can give information on the residual stress state of the sample. Moreover, the intensity of the above mentioned bands was used to non-destructively investigate the surface finishing of the acetabular cups, according to other authors (Garcia et al., 2002).

The fluorescence spectra were obtained using an argon–krypton laser (Innova Coherent 70) operating at 488 nm to excite the fluorescence and a Jasco NRS-2000C micro-Raman spectrometer equipped with a 160 K frozen digital CCD detector (Spec-10: 100B, Roper Scientific Inc.) to collect the excited fluorescence. To ensure that no laser heating occurred and contributed to the observed frequency shifts, all measurements were performed at a low laser power (i.e. 1 mW). Instrumental fluctuations represent another source of possible variation in the measured frequency. In order to correct for this, a characteristic neon line at 14 431 cm$^{-1}$ was used as a frequency calibration standard.

The spectra were recorded in back-scattering conditions with 1 cm$^{-1}$ spectral resolution using an objective lens of 10× magnification; the laser spot size was larger than the grain size of the ceramics, assuring that the fluorescence was being averaged over a large number of grains. Moreover, to obtain a good representation of the stress distribution, ten spectra were collected in ten different points of each sample.

Ten spectra were recorded, on each sample, in the inner surface near the centre (in a spatial range of about 1.5 mm from the centre). A soaked unworn cup of each set of specimens was analyzed as control.

The bands monitored were at about 14 396 cm$^{-1}$ ($R_1$) and 14 424 cm$^{-1}$ ($R_2$). Their width (expressed as full width at half maximum, FWHM), intensity and frequency were determined by fitting the experimental spectra with mixtures of Lorentzian and Gaussian functions. The fitting was done using a commercial software (OPUS 5.0, Bruker Optik GmbH, Germany).

## 2.5. Crack resistance measurements

Short-crack resistance curves (*R*-curve behaviour) were measured and analyzed by the indentation-strength in the bending (ISB) method (Braun et al., 1992) using prismatic bars in which the centres of the tensile faces, polished down to 1 μm, were indented with a Vickers diamond at contact loads, *P*, between 10 and 500 N. The specimens were tested at room temperature using a three-point support with a span of 40 mm in an universal testing machine (Instron Model 4411). The specimens were loaded to failure with a cross-head speed of 0.05 mm/min. The mechanical test was performed immediately after indentation to avoid any subcritical crack growth due to stress corrosion effects. Special effort was made to examine all specimens after testing using reflected light optical microscopy (Leica, DMR model), to verify that the indentation contact site acted as the origin of failure. *R*-curves were measured considering a radial crack c produced by the indentation at a load *P* and subjected to the action of a tensile stress *σa* due to the applied stress load during three-point bending. During post-indentation bending, the crack is subjected to a total stress intensity, $K_t$, which is the sum of contributions from the residual stress intensity factor acting on the indentation crack resulting from the elastic–plastic mismatch associated with the indentation, $K_r$, and the stress intensity factor resulting from the applied stress, $K_a$:

$$K_t(c) = K_a(c) + K_r(c) = \psi \cdot \sigma a \cdot c^{1/2} + \xi \cdot (E/H)^{1/2} \cdot P/c^{3/2} = K_R(c) \quad \text{(eq. 1)}$$

where $\psi$ is a crack geometry factor, $\xi$ is the dimensionless geometrical constant, *E* is Young's modulus, *H* hardness, and $K_R$ is the crack resistance of the material. For a given indentation load, *P*, failure is assumed to occur at the stress where the applied stress $\sigma_a$ is equal to the fracture stress, $\sigma_f$, which satisfies the following balance and tangency conditions:

$$K_t(c) = K_R(c) \quad \text{(eq. 2)}$$

$$dK_t(c)/dc = dK_R(c)/dc \quad \text{(eq. 3)}$$

The *R*-curve is determined by solving Eqs. (2) and (3) for each beam simultaneously.

## 2.6. Crack velocity functions and threshold determination

Crack velocities from $10^{-12}$ m s$^{-1}$ (necessary for threshold determination) to $10^{-2}$ m s$^{-1}$ (fast fracture) were measured by the double torsion technique, in order to get insight on the crack velocity versus stress intensity factor functions. The double torsion specimens (40 mm∗20 mm∗2 mm plates) and the loading configuration are shown in Fig. 2. The tensile surface is polished down to 1 mm in order to observe the crack with a precision of ±2 mm. A notch of dimension $a_o$=10mm and root $\rho$=0.1mm is machined with a diamond saw and an indentation performed at low load (5 kg) in order to initiate a small crack. Subsequent pre-cracking is performed by loading the specimens at low rate in order to induce a 'real' sharp crack of initial length $a_i$=12mm. The double torsion configuration has for a long time been known to give rise to a stress intensity factor which is independent of crack length, given by (Shyam and Lara-Curzio, 2006):

$$K_I = \frac{W_m}{U^2}\left(\frac{3(1+\nu)}{W\psi}\right)^{1/2} \qquad (eq.\ 4)$$

$P$ is the load, $W_m$ the span, $U$ and $W$ the width and the thickness of the specimen, $\nu$ the Poisson's ratio (taken here as equal to 0.3), and $\psi$ a calibration factor.

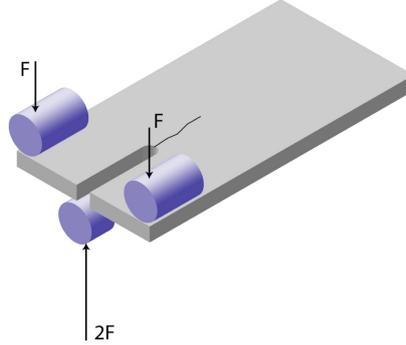

Fig. 2. Double Torsion specimen and loading configuration.

However, it has been demonstrated recently, both experimentally and by numerical simulation (Chevalier et al., 1996; Ciccotti, 2000) that $K_I$ is slightly dependent on the crack length. To obtain accurate $V$–$K_I$ diagrams a correction factor should be introduced in the conventional expression of $K_I$, as expressed with the following empirical equation:

$$K_I = HP\left(\frac{a}{a_0}\right)^\gamma \qquad (eq.5)$$

$a_0$ is the notch length and $a$ is the total crack length. $\gamma$ depends on the geometry of the test sample. For the dimensions used for the present study $\gamma$=0.19 (Chevalier et al., 1996).

Crack velocity functions were determined via two methods: relaxation tests and constant-loading tests. The load-relaxation method, which was first reported by Williams and Evans (1973), was used to obtain the slow crack growth $V$–$K_I$ diagrams in the velocity range $10^{-2}$–$10^{-7}$ m/s. This method does not allow measurements at very low velocity but it does present the advantage of being quick to obtain measurements at high velocities. Pre-cracked specimens are loaded at a constant rate of 0.2 mm/min, followed by subsequent stopping of the cross-head at constant displacement, when the crack starts to propagate. The obtained load-relaxation versus time ($P$ versus $t$) plot allows the determination of the $V$–$K_I$ curve by a compliance calibration (Chevalier et al., 1996; Ciccotti, 2000; Shyam and Lara-Curzio, 2006; Williams and Evans, 1973).

Measurement of the crack velocities $V$ under constant load presents the advantage of allowing the measurement of very low velocities, down to $10^{-12}$ m/s. Thus, the specimens are subjected to different static loads under a prescribed duration $\Delta t$. The crack length is measured via optical microscopy, with a precision of ±2 µm, and $V$ is defined as the ratio of crack increment $\Delta a$ to the duration $\Delta t$:

$$V = \frac{\Delta a}{\Delta T} \qquad \text{(eq. 6)}$$

Crack velocities were also measured under cyclic loading, at a frequency of 1 Hz on the same testing machine than for static fatigue measurements, which presents the advantage of allowing a direct crack velocity comparison with static tests. The specimens were loaded at an imposed sine shape load between $P_{\min}$ and $P_{\max}$, for a given duration $\Delta t$. The $R = K_{I\min}/K_{I\max} = P_{\min}/P_{\max}$ ratio was selected to 0.1. The crack length increment $\Delta a$ was again measured by optical microscopy with a precision of ±2 mm and the crack velocity $V$ by Eq. (3).

## 2.7. Aging resistance

The transformation being both thermally activated and accelerated by the presence of water, samples were put in an autoclave in steam during controlled duration at 134 °C, under 2 bars pressure, in order to induce the phase transformation at the surface with time. Knowing the thermal activation (−106 kJ/mol) of the aging process (Chevalier et al., 1999a), it is possible to calculate that 1 h of such a treatment would correspond roughly to three years *in vivo*. The transformation was followed by measuring the monoclinic phase fraction evolution by X-ray diffraction (XRD). XRD data were collected with a *θ-2θ* diffractometer using the Cu-K$_\alpha$ radiation. Diffractograms were obtained from 27° to 33°, at a scan speed of 0.2 °/min and a step size of 0.02°. The monoclinic phase fraction $X_m$ was calculated using the Garvie and Nicholson method (Garvie and Nicholson, 1972), modified by Toraya et al. (1984):

$$X_m = \frac{I_m(-111)+I_m(111)}{I_m(-111)+I_m(111)+I_t(101)} \qquad \text{(eq. 7)}$$

$I_t$ and $I_m$ represent the integrated intensity (area under the peaks) of the tetragonal (101) and monoclinic (111) and (−111) peaks. The monoclinic volume fraction, $V_m$, is then given by:

$$V_m = \frac{1.311 X_m}{1+1.311 X_m} \qquad \text{(eq. 8)}$$

## 2.8. Fracture resistance of femoral heads prototypes

Compressive loads to failure of NK heads were performed according to ISO-DIS 7206-10 protocol. In order to evaluate the potential of the nano-composite for critical designs, 28 mm long neck femoral heads were coupled with CrCo conical tapers. 7 heads were loaded in compression (crosshead speed: 0.5 mm/min) according to the configuration of Fig. 3, on a universal INSTRON hydraulic testing machine. 7 AL femoral heads were tested with the same testing configuration.

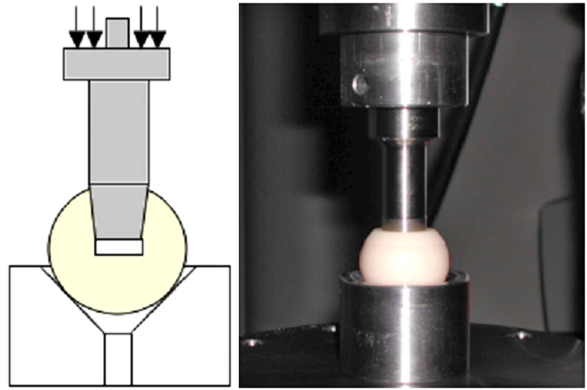

Fig. 3. Testing configuration for the measurement of the fracture resistance of alumina–zirconia nanocomposite prototype femoral heads.

## 3. Results and discussion

### 3.1. Inspection of worn ceramic components after 7 million cycles

Strain can induce a shift of the characteristic frequencies of spectral bands, i.e. Raman, infrared or luminescence bands. However, in this work we concentrated on the $R_1$ and $R_2$ fluorescence bands as a function of stress. The reason for this is that the fluorescence signal is several orders of magnitude greater in intensity than the Raman bands (Fig. 4) and hence more precise measurements can be conducted.

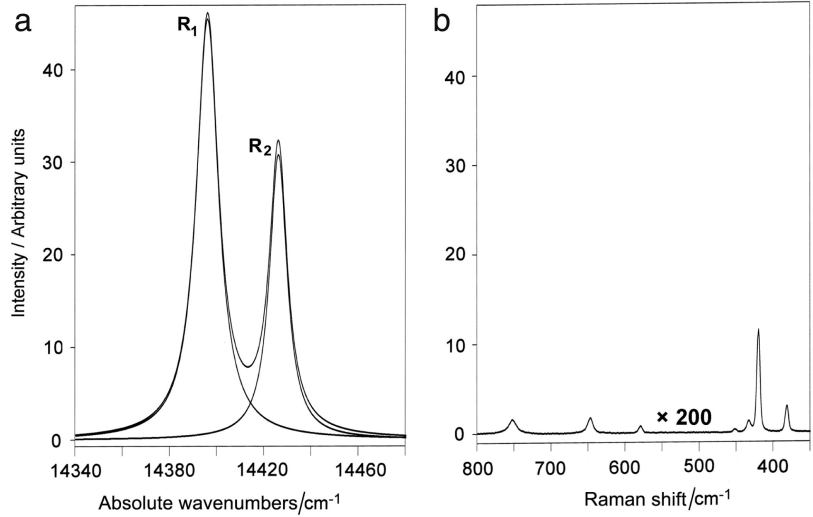

Fig. 4. Photoluminescence (a) and Raman (b) spectra of a control unworn commercial alumina acetabular cup. The $R_1$ and $R_2$ fluorescence bands (a) are several orders of magnitude greater in intensity than the Raman spectrum. The fluorescence spectrum was fitted into the two $R_1$ and $R_2$ components to more precisely evaluate band frequencies, intensities and FWHM.

All the samples contained an adequate $Cr^{3+}$ impurity level for the $R_1$ and $R_2$ bands to be recorded with a high signal-to-noise ratio, so that precise measurements of band frequency, intensity and FWHM were assured. As an example, Fig. 4(a) reports the fluorescence spectrum of a control unworn commercial alumina acetabular cup fitted into the two $R_1$ and $R_2$ components.

The data obtained from the fitting of the photoluminescence spectra are reported in Table 1. By comparing the three sets of control unworn specimens, the most significant observable change involved the intensity of both $R_1$ and $R_2$ bands; less significant changes were observed for their frequencies and FWHM. As regards the latter, its value has been reported to be correlated with grain size and microcracking. Actually, microcracks are more easily formed in materials with large grain size (Rice and Pohanka, 1979) and are known to reduce the width of the Gaussian residual stress distribution and thus the FWHM of the fluorescence bands (Ortiz and Suresh, 1993).

Table 1. Frequency, intensity and full width at half maximum (FWHM) of the $R1$ and $R2$ bands (mean values ± standard deviations) as obtained by fitting the experimental.
[a] Mean values obtained on three worn cups.

| Sample | | Frequency (± standard deviation) | $R1$ band intensity (± standard deviation) | FWHM (± standard deviation) | Frequency (± standard deviation) | $R2$ band intensity (± standard deviation) | FWHM (± standard deviation) |
|---|---|---|---|---|---|---|---|
| Commercial alumina | Control | 14 396.2 ± 0.1 | 45 ± 2 | 11.63 ± 0.05 | 14 424.2 ± 0.1 | 30 ± 1 | 9.59 ± 0.02 |
| | Worn [a] | 14 396.1 ± 0.1 | 45 ± 2 | 11.57 ± 0.08 | 14 424.1 ± 0.1 | 30 ± 1 | 9.52 ± 0.04 |
| Experimental alumina | Control | 14 396.0 ± 0.1 | 18 ± 2 | 11.52 ± 0.02 | 14 424.0 ± 0.1 | 12 ± 1 | 9.31 ± 0.01 |
| | Worn [a] | 14 396.1 ± 0.2 | 17 ± 2 | 11.50 ± 0.04 | 14 424.2 ± 0.1 | 12 ± 1 | 9.31 ± 0.09 |
| experimental NZTA | Control | 14 395.8 ± 0.1 | 25 ± 2 | 11.39 ± 0.01 | 14 423.7 ± 0.1 | 16 ± 1 | 9.35 ± 0.01 |
| | Worn [a] | 14 395.8 ± 0.1 | 25 ± 2 | 11.38 ± 0.06 | 14 423.8 ± 0.1 | 16 ± 1 | 9.31 ± 0.06 |

As can be easily seen from Table 1, the intensity of the fluorescence bands for the unworn control samples increased along the series: BK<NK<AL.

Recent studies have indicated the dependence of photoluminescence on surface quality (Garcia et al., 2002). It can be affirmed that the higher the roughness of the sample, the lower the photoluminescence intensity. In fact, pores or scratches act as scattering centers of the incident laser beam and reduce the transmission and hence excitation depth. In this light, it can be affirmed that commercial alumina showed the best surface finishing. Going from this set of samples to NK the surface quality worsened and the worsening was even more pronounced for experimental alumina. However, it should be remembered that the analyzed NK and experimental alumina specimens were only prototypes and therefore their surface finishing can be worse than for a production type, as previously observed for ceramic couplings (Affatato et al., 2001). Hip joint wear simulator tests did not significantly alter the surface finishing and residual stress state of the three sets of acetabular cups; no significant changes in frequency and FWHM of the $R_1$ and $R_2$ bands were observed upon wear testing (Table 1). However, it must be recalled that from a statistical point of view, the three sets of specimens did not show significant differences in wear behaviour in hip joint wear simulator tests (Affatato et al., 2006). Interestingly, gravimetric measurements showed the same trend as photoluminescence intensities: the highest and lowest weight losses were observed for AL and BK samples, respectively, while the NK specimens were characterized by intermediate weight losses (Affatato et al., 2006).

These findings suggest that the better the sample finishing, the better the wear behaviour. Therefore, full density should be reached to limit wear in ceramics. This is particularly true for the cups, where it is hard to obtain excellent forming conditions and surface finishing (concave surface).

Scanning Electron Microscopy (Fig. 5) and Atomic Force Microscopy (Table 2) observations confirm the fact the commercial alumina exhibits the best initial surface finish, followed by NZTA and experimental alumina prototypes. This leads then to less surface damage after 7 million cycles. Even if the damage is low for all ceramics of the study, it can be argued that the experimental alumina exhibits the most important amount of grain pull out and third body wear. We must remind however, the amount of wear debris generated by such ceramic–ceramic configurations is exceptionally low when compared to standard metal–polyethylene or even ceramic-polyethylene.

Table 2. $R_a$ (roughness) values measured at the top of a commercial alumina head (AL), an experimental alumina head (BK) and a NK head, before and after 7 million cycles.

|  | Commercial alumina (AL) | Experimental alumina (BK) | Experimental nano-composite (NK) |
| --- | --- | --- | --- |
| Ra (0 Mc) | 2.3 (nm) | 7.1 (nm) | 5.1 (nm) |
| Ra (7 Mc) | 3.0 (nm) | 18 (nm) | 11 (nm) |

## 3.2. Crack resistance measurements

The fracture behaviour of short cracks has been studied in flexure using the indentation method to produce controlled surface cracks. The advantage of this technique is that it is able to provide an assessment of the mechanical properties at the correct microstructural length scale. The *R*-curve behaviour is caused by the increase of fracture toughness with increasing crack length due to crack-tip process zone phenomena and/or crack bridging due to interlocking grains. For example, in a coarse grained alumina (~16 μm grain size) the fracture toughness saturates at ~6 MPa m$^{1/2}$, which is about double its short crack value (Reichl and Steinbrench, 1988). Whilst this is an impressive increase in fracture toughness in a material, it should be realised that the critical flaws in these experiments were much larger (>1 mm) than would be encountered in real components. On the other hand, for short cracks in the range of 50–200 μm, the effect of these toughening mechanisms can be significantly reduced. Therefore, in applications where small cracks are of interest, i.e., in biomedical implants, *R*-curves measured on long cracks are not relevant. On the contrary, the increase in the initial fracture toughness and the crack growth resistance in the short crack region (<100 μm) is necessary. Moreover, the fracture toughness operative at small crack-size scales is a relevant material property which controls deformation response during wear (Scattergood et al., 1991).

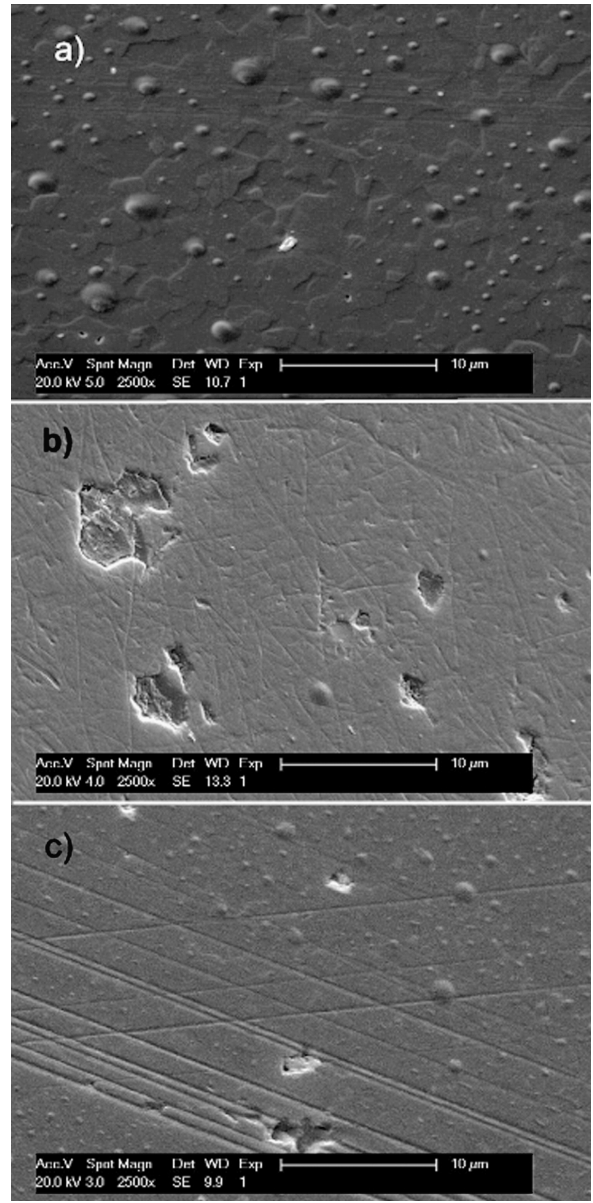

Fig. 5. Scanning Electron Microscopy images of ceramic heads after 7 millions cycles hip simulator study. (a) Commercial alumina (AL), (b) experimental alumina (BK), (c) experimental alumina–zirconia nano-composite (NK).

The resulting short-crack resistance curves obtained for alumina, zirconia, zirconia toughened alumina "micro"-composite and nanostructured alumina–zirconia composite have been plotted in Fig. 6. The result for the alumina shows no such rising *R*-curve. For monolithic alumina, the main toughening mechanism was grain bridging by large elongated $Al_2O_3$ grains behind the crack tip. Due to the small grain size of the biomedical grade alumina, no bridging effect would be expected.

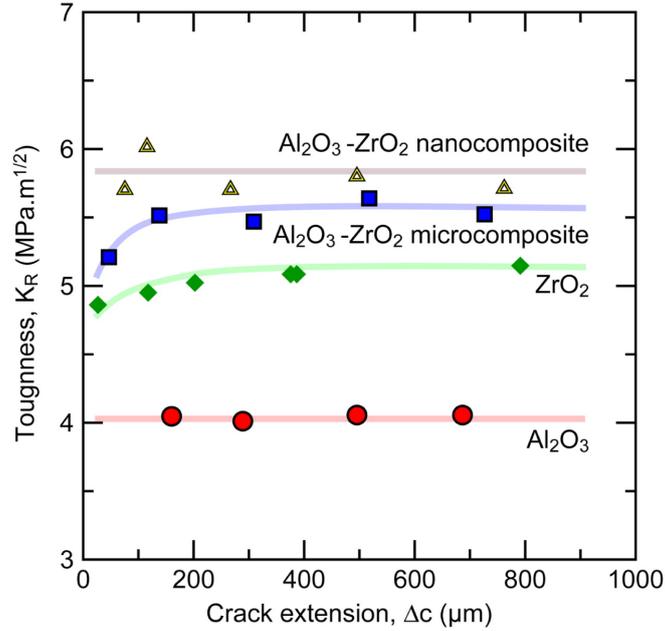

Fig. 6. Short-crack resistance curves obtained for biomedical grade alumina, yttria stabilized zirconia, conventional zirconia toughened alumina 'micro'-composite and currently developed nano-structured alumina–zirconia composite.

The alumina–zirconia nanocomposite specimens show a flat *R*-curve with the highest toughness value. In these nanocomposites, the toughening mechanisms operate on a scale smaller than that of the matrix microstructure, enhancing the "intrinsic" fracture properties of the material.

On the other hand, zirconia and alumina–zirconia "micro"-composites show a slight rising crack growth resistance with crack extension due to a transformation toughening mechanism that promote a process zone in the neighbourhood of the crack tip. Consistent with the very fine microstructure and the relatively narrow transformation zone of these ceramics, the *R*-curve rises steeply during the first 100 μm extension of the crack. Over this crack length, fracture resistance reaches a plateau toughness, meaning that toughness no longer increases with crack extension.

### 3.3. Crack velocity functions and threshold determination

Fig. 7 shows crack velocity diagrams under static loading, for the nanostructured alumina–zirconia composite, standard biomedical grade alumina and zirconia ceramics, and for a zirconia toughened alumina 'micro'-composite developed previously. The results for the four ceramics show the typical three stages of Slow Crack Growth (SCG) and a threshold, below which no crack propagation occurs. The threshold ($K_{I0}$), was determined from the points on the $V$–$K_I$ diagram, below which there is an abrupt drop of the crack velocity, $V<10-12 ms^{-1}$. On the other hand, the toughness ($K_{IC}$) was determined by extrapolation of the $V-K_I$ curve to high crack velocities ($10^{-2}$ m s$^{-1}$). Their values are represented in Table 3 for the different ceramics.

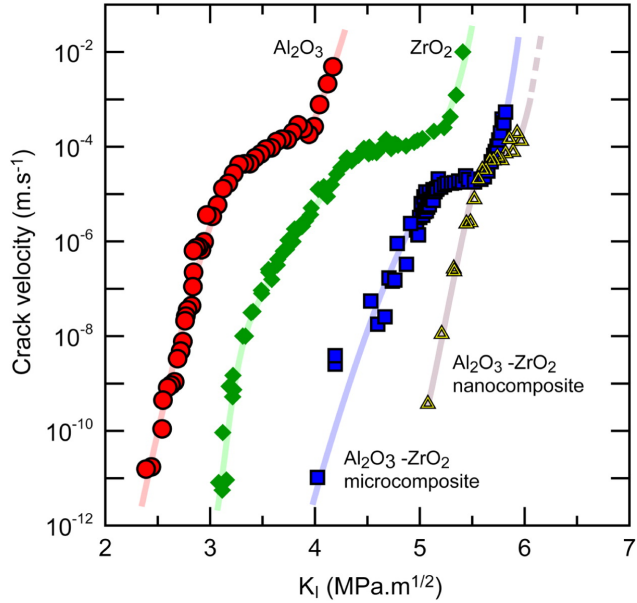

Fig. 7. Crack velocity diagrams under static loading, for biomedical grade alumina, yttria stabilized zirconia, conventional zirconia toughened alumina 'micro'-composite and currently developed nano-structured alumina–zirconia composite.

Table 3. Toughness ($K_{IC}$), static fatigue threshold ($K_{I0s}$) and cyclic fatigue threshold ($K_{I0c}$) of the currently developed alumina–zirconia nanocomposite (NK), alumina–zirconia micro-composite (ZTA), biomedical grade alumina (AL) and yttria-stabilized zirconia (3Y-TZP).

|  | Alumina–zirconia nano-composite (NK) | Alumina–zirconia micro-composite (ZTA) | Biomedical grade alumina (AL) | Biomedical grade alumina (3Y-TZP) |
|---|---|---|---|---|
| $K_{IC}$ (MPa√m) | 6.2 ± 0.2 | 6.0 ± 0.2 | 4.2 ± 0.2 | 5.5 ± 0.2 |
| $K_{I0S}$ (MPa√m) | 5.0 ± 0.2 | 4.0 ± 0.2 | 2.5 ± 0.2 | 3.2 ± 0.2 |
| $K_{I0C}$ (MPa√m) | 5.0 ± 0.2 | Not measured | 2.5 ± 0.2 | 2.8 ± 0.2 |

Contrasting results are first underlined for alumina and zirconia monolithic ceramics. Zirconia exhibits a higher toughness than alumina but their thresholds are close, meaning that the necessary crack tip stress to initiate slow crack growth is roughly the same in both materials. The high toughness of zirconia for a monolithic ceramic is attributed to the stress induced phase transformation of metastable tetragonal grains towards the monoclinic symmetry ahead of a propagating crack, leading to an increase of the work of fracture (Evans and Heuer, 1980). This phenomenon is referred to as 'transformation toughening'. Alumina has lower susceptibility to water and thus to stress assisted corrosion. As a consequence, the $V$–$K_I$ curve of alumina presents a higher slope than the curve corresponding to zirconia. Even if alumina is intrinsically more brittle than zirconia (lower toughness), it exhibits a threshold of the same order. From an atomistic point of view, this means that the fracture energy of zirconia is lower in the presence of water or body fluid, because the zirconia bonds are prone to chemisorption of the polar water molecules (like silica glass for example). In practical terms, the benefit of using zirconia instead of alumina is limited if we consider long-term behaviour.

The zirconia-toughened alumina 'micro'-composite possesses both a larger toughness and threshold than the two monolithic ceramics. First, alumina rich composites present crack propagation through the alumina matrix. Thus, these composites possess a lower susceptibility to stress assisted corrosion by water or body fluid. Second, these materials are reinforced by the presence of transformable zirconia particles, shifting the $V$–$K_I$ diagram of alumina towards higher $K_I$ values. The authors have shown (De Aza et al., 2002) in a previous work that the presence of small amounts of transformable zirconia particles in a given matrix leads to a shift of the $V$–$K_I$ diagram towards higher $K_I$ values, preserving the slope of the curve. Therefore, the large slope of the $V$–$K_I$ diagram of alumina is preserved, but the diagram is shifted to large $K_I$ values due to transformation toughening, which means in turn a large toughness and threshold. It is worth noting that 10 vol.% of transformable zirconia particles, with a size ranging between 100 and 600 nm, corresponds to the maximum amount of transformation toughening (De Aza et al., 2003).

Including a smaller amount of zirconia particles, with a size lower than 100 nm totally hinders the possibility of transformation toughening, since the particles become too small to transform even at the crack tip. This was verified by X-ray Diffraction, which revealed no transformation from the tetragonal to the monoclinic symmetry at the surface of fractured nanostructured alumina–zirconia composites. However, both the toughness and threshold of the alumina–zirconia composite are significantly larger than that of the 'micro'-composite, and much larger than alumina. This means that only 1.7 vol.% of zirconia nano-particles in the alumina matrix dramatically improve its crack resistance. This dramatic increase of SCG resistance can be attributed neither to transformation toughening, nor crack bridging, since the $R$-Curve was proven to be flat (Fig. 6). Moreover, the mode of failure was predominantly transgranular. On some occasions, some crack bridging ligaments were observed on the crack path but their number was clearly too low to account for a significant toughening effect. Including transgranular nano-particles of a second phase with a different thermal expansion coefficient, however, induces large residual stresses in the composite, which may have a strong impact on the SCG behaviour. In order to determine the presence of a residual stress field in the alumina matrix, high angular precision diffraction patterns (monocromatized incident beam Philips Xpert diffractometer) were recorded between 27° and 45° for pure sintered alumina and for the nano-composite. Small angular displacements were found in the α-alumina peaks of nanocomposite corresponding to the following planes: (104), (110), (006), (113). According to these data, a compressive strain of $3 \times 10^{-4}$ and $2 \times 10^{-4}$ was found for the *a* and *c* axes of the α-alumina matrix. These strains correspond to a compressive average stress of $150 \pm 50$ MPa. These compressive stresses superimpose on the stresses applied by the external stress field. In other words, the stress intensity factor at the crack tip is lowered in the presence of the residual compressive stresses. For a semi-elliptical flaw of dimension a subjected to a uniform tensile stress, the actual stress intensity factor at the crack tip is given by:

$$K_{I\ tip} = (\sigma_a - \sigma_{res})\sqrt{\pi a} \qquad \text{(eq. 9)}$$

where $\sigma_a$ is the applied stress, and $\sigma_{res}$ the residual compressive stress. The ratio between $K_{I\ tip}$ (stress intensity factor at the crack tip) and $K_I$ (the applied stress intensity factor) is therefore given simply by:

$$\frac{K_{I\ tip}}{K_I} = \frac{(\sigma_a - \sigma_{res})}{\sigma_a} \qquad \text{(eq. 10)}$$

The relative influence of the residual stress field is therefore higher for low applied stress intensity factor (i.e. around $K_{I0}$) than for high applied $K_I$ (i.e. at $K_{IC}$), which is in agreement with a higher slope of the apparent $V$–$K_I$ diagram. In other words, transformation toughening in micro-composites leads to a shift of the $V$–$K_I$ curve, preserving the slope of the alumina matrix, while residual stresses in nanostructured, intra-type zirconia particles, composites lead to an increase of the $V$–$K_I$ slope.

The comparative sensitivity of ceramics to SCG can be plotted in a normalised $V$–$K_I/K_{IC}$ diagram, where $K_{IC}$ is the toughness. The higher the slope of the diagram, i.e. the higher the $K_{I0}/K_{IC}$ ratio, the lower the sensitivity to SCG by stress assisted corrosion. Fig. 8 represents a schematic summary of results obtained in the different materials of the study and compared to covalent ceramics (SiC and $Si_3N_4$). The results illustrate the commonly accepted idea that the higher the covalent to ionic bonding ratio, the lower the susceptibility to SCG. This is directly related to the atomic structure of the material. The nano-composite exhibits a peculiar behaviour, with a slope of the $V$–$K_I/K_{IC}$ diagram and a $K_{I0}/K_{IC}$ ratio close to covalent ceramics (the $Al_2O_3$–$nZrO_2$ lying between $Si_3N_4$ and SiC).

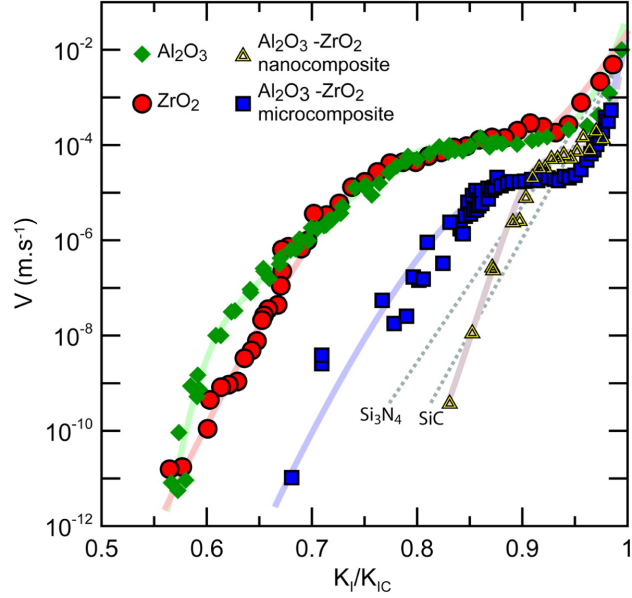

Fig. 8. $V$–$K_I/K_{IC}$ laws of biomedical grade alumina, yttria stabilized zirconia, conventional zirconia toughened alumina 'micro'-composite and currently developed nanostructured alumina–zirconia composite. Schematic $V$–$K_I/K_{IC}$ laws of covalent ceramics (SiC and $Si_3N_4$) are given for comparison.

In order to evaluate the cyclic fatigue behaviour of the alumina–zirconia nano-composite, crack velocities were also measured under cyclic loading. The tests were conducted at a frequency of 1 Hz on the same testing machine as for static fatigue measurements, which presents the advantage of allowing a direct crack velocity comparison with static tests. The specimens were loaded at an imposed sine shape load between $P_{min}$ and $P_{max}$, for a given duration $\Delta t$. The $R = K_{Imin}/K_{Imax} = P_{min}/P_{max}$ ratio was selected to be 0.1. The crack velocity function under cyclic fatigue is compared to that obtained under static fatigue in Fig. 9. The comparison of the thresholds for the different materials, under cyclic and static fatigue, is given in Table 3. A slight increase of crack velocity and a tendency to a lower threshold under cyclic loading is noticed. However, the cyclic fatigue degradation is low, in particular in comparison to monolithic zirconia or alumina–zirconia micro-composites, where transformation toughening acts to resist crack propagation but is degraded under cyclic loading (Chevalier et al., 1999b,c). Pure alumina is not affected by cyclic loading since the amount of crack bridging is negligible for that grain size (Attaoui et al., 2005). Since residual stresses are elastic by nature, the only source of cyclic degradation in the alumina–zirconia nanocomposite is the small amount of crack bridging observed under static fatigue. However, the threshold below which no crack propagation occurs, still stands well above that of the other ceramic materials, providing a large crack resistance for long-term applications.

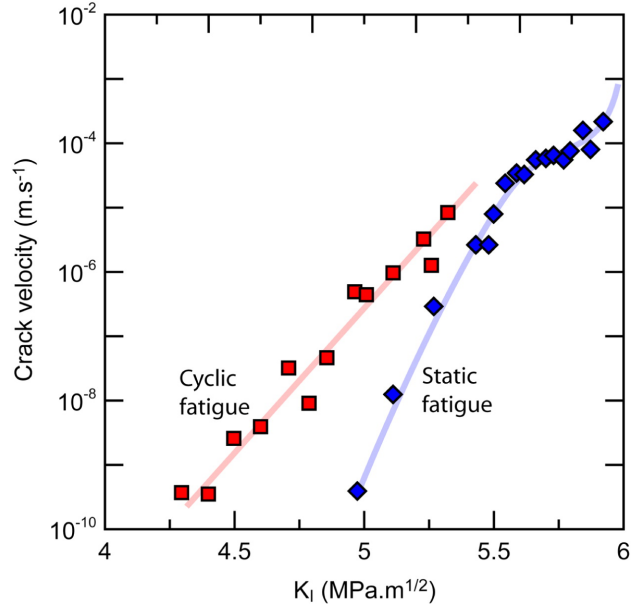

Fig. 9. Crack velocity function of the nano-structured alumina–zirconia composite under cyclic fatigue, compared to that obtained under static fatigue.

## 3.4. Aging resistance

Fig. 10 represents the aging kinetics of the 3Y-TZP, alumina–zirconia micro-composite and nano-composite, in terms of monoclinic fraction increase versus time. As expected, the 3Y-TZP ceramic undergoes rapid surface transformation, since the monoclinic fraction reaches saturation after less than 10 h in the autoclave (i.e. roughly 30 years *in vivo*). On the contrary, the two alumina–zirconia composites do not exhibit any phase transformation during aging, even for long (un-realistic) duration. Previous papers published recently on the aging behaviour of alumina–zirconia composites have shown that alumina–zirconia could be considered as safe against aging, provided that the zirconia content in the composite is kept lower than 10% in the case of un-stabilized zirconia and that no aggregates are present in the composite. The present results stand on this line. This ensures perfect stability of the alumina–zirconia composite, which is a clear advantage versus 3Y-TZP zirconia.

## 3.5. Fracture resistance of femoral heads

The load to failure of 7 prototypes processed from the alumina–zirconia nanocomposite is given in Table 4. The results show that, even for the critical design chosen for the load to failure test (28 mm heads, long neck, against CoCr taper), the prototypes satisfy the ISO 7206-10 standard:

- The mean load to failure (49.2 kN) is higher than 46 kN,
- The lowest load to failure reported is higher than 20 kN.

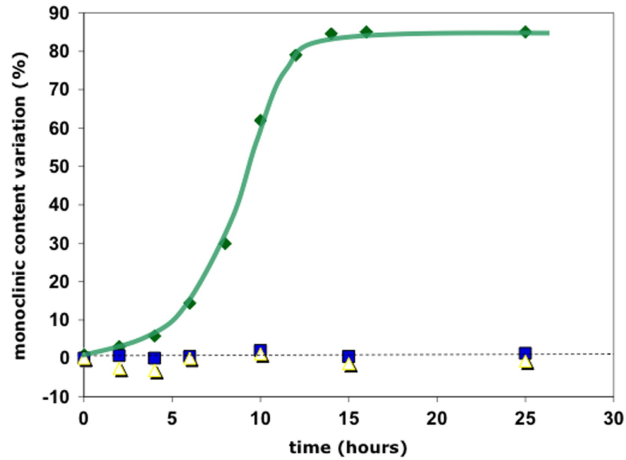

Fig. 10. Aging kinetics (monoclinic fraction increase versus time at 134°C, 2 bars) of biomedical grade zirconia (3Y-TZP), alumina–zirconia micro-composite and nano-composite.

Table 4. Load to failure measured on 7 femoral heads prototypes processed from alumina–zirconia nano-composites.

| Femoral head n° | Load to failure (kN) |
|---|---|
| 1 | 50.96 |
| 2 | 34.50 |
| 3 | 54.06 |
| 4 | 44.96 |
| 5 | 58.82 |
| 6 | 45.75 |
| 7 | 55.56 |
| **Mean value** | **49.23** |

For the same design and the same testing configuration, the commercial alumina failed to satisfy the ISO requirements. Indeed, the main load to failure (39 kN) was below the required value.

## 4. Conclusions

Due to the modest failure resistance of alumina and the problem of yttria stabilized zirconia in terms of long term reliability, there is a trend today to develop new alternatives, especially for critical and/or new designs for which alumina does not satisfy mechanical requirements. The aim of the present paper was to present a global picture of the mechanical performance and the durability of newly developed alumina–zirconia nanocomposites. We have shown that such nanocomposites exhibit a very limited surface damage after 7 million cycles in a hip simulator. They also exhibit a crack resistance under static and cyclic fatigue well beyond that of all existing biomedical grade ceramics, opening the possibility to develop critical designs that are not possible with monolithic alumina. Associated to an expected full stability in vivo, the overall set of results ensures a potential future development for these kinds of new nanocomposites in the orthopedic field.

## Disclosure statement

All authors disclose any actual or potential conflict of interest including any financial, personal or other relationships with other people or organizations within three (3) years of beginning the work submitted that could inappropriately influence (bias) their work.

## Acknowledgement

The work was funded by the EU under contract $n°$G5RD-CT-2001-00483 (BIOKER).

## References

Affatato et al., 2001: S. Affatato, M. Goldoni, M. Testoni, A. Toni, Mixed oxides prosthetic ceramic ball heads. Part 3: effect of the ZrO2 fraction on the wear of ceramic on ceramic hip joint prostheses. A long-term in vitro study, Biomaterials, 22 (2001), pp. 717-723

Affatato et al., 2006: S. Affatato, R. Torrecillas, P. Taddei, M. Rocchi, C. Fagnano, G. Ciapetti, A. Toni, Advanced nanocomposite materials for orthopaedic applications. I. A long-term in vitro wear study of zirconia-toughened alumina, J. Biomed. Mater. Res. Appl. Biomater., 78 (2006), pp. 76-82

Attaoui et al., 2005 : H.E. Attaoui, M. Saâdaoui, J. Chevalier, G. Fantozzi, Quantitative analysis of crack shielding degradation during cyclic fatigue of alumina, J. Am. Ceram. Soc., 88 (2005), pp. 172-178

Braun et al., 1992 : L.M. Braun, S.J. Benninson, B.R. Lawn,Objetive evaluation of short-crack toughness curve using indentation flaws: case study on alumina-based ceramics, J. Am. Ceram. Soc., 75 (1992), pp. 3049-3057

Cales, 2000: B. Cales, Zirconia as a sliding material: histologic, laboratory, and clinical data, Clin. Orthop., 379 (2000), pp. 94-112

Campbell et al., 2004: P. Campbell, F. Shen, H. McKellop, Biologic and tribologic consideration of alternative bearing surfaces, Clin. Orthop., 418 (2004), pp. 98-111

Chevalier, 2006: J. Chevalier, What future for zirconia as a biomaterial? Biomaterials, 27 (2006), pp. 535-543

Chevalier et al., 1999a : J. Chevalier, B. Cales, J.M. Drouin, Low temperature aging of Y-TZP ceramics, J. Am. Ceram. Soc., 82 (1999), pp. 2150-2154

Chevalier et al., 2000: J. Chevalier, A. De Aza, M. Schehl, R. Torrecillas, G. Fantozzi, Extending the lifetime of ceramic orthopedic implants, Adv. Mater., 12 (2000), pp. 1619-1621

Chevalier et al., 2005: J. Chevalier, S. Deville, G. Fantozzi, J.F. Bartolomé, C. Pecharroman, J.S. Moya, L.A. Diaz, R. Torrecillas, Nanostructured ceramic oxides with a slow crack growth resistance close to covalent materials, Nanoletters, 5 (2005), pp. 1297-1301

Chevalier et al., 2007 : J. Chevalier, L. Gremillard, S. Deville, Low temperature degradation of zirconia and implications on biomedical implants, Ann. Rev. Mater. Res., 37 (2007), pp. 1-32

Chevalier et al., 1999b : J. Chevalier, C. Olagnon, G. FantozziCrack propagation and fatigue in zirconia-based composites

Composites, Part A, 30 (1999), pp. 525-530

Chevalier et al., 1999c : J. Chevalier, C. Olagnon, G. Fantozzi, Subcritical crack propagation in 3Y-TZP ceramics: static and cyclic fatigue, J. Am. Ceram. Soc., 82 (1999), pp. 3129-3138

Chevalier et al., 1996 : J. Chevalier, M. Saadaoui, C. Olagnon, G. Fantozzi, Double torsion testing a 3Y-TZP ceramic, Ceram. Int., 22 (1996), p. 171

Ciccotti, 2000: M. Ciccotti, Realistic finite-element model for the double-torsion loading configuration, J. Am. Ceram. Soc., 83 (2000), pp. 2737-2744

Dauskardt et al., 1994: R.H. Dauskardt, R.O. Ritchie, J.K. Takemoto, A.M. Brendzel, Cyclic fatigue and fracture in pyrolytic carbon-coated graphite mechanical heart-valve prostheses: role of small cracks in life prediction, J. Biomed. Mater. Res. Appl. Biomater., 28 (1994), pp. 791-804

De Aza et al., 2002: A.H. De Aza, J. Chevalier, G. Fantozzi, M. Schehl, R. Torrecillas, Crack growth resistance of alumina, zirconia and zirconia toughened alumina ceramics for joint prostheses, Biomaterials, 23 (2002), pp. 937-945

De Aza et al., 2003 : A.H. De Aza, J. Chevalier, G. Fantozzi, M. Schehl, R. Torrecillas, Slow-crack-growth behavior of zirconia-toughened alumina ceramics processed by different methods, J. Am. Ceram. Soc., 86 (2003), pp. 115-120

Deville et al., 2003: S. Deville, J. Chevalier, G. Fantozzi, J.F. Bartolomé, J. Requena, J.S. Moya, R. Torrecillas, L.A. Diaz, Low-temperature ageing of zirconia-toughened alumina ceramics and its implication in biomedical implants, J. Eur. Ceram. Soc., 23 (2003), pp. 2975-2982

Evans and Heuer, 1980: A.G. Evans, A.H. Heuer, Review-transformation toughening in ceramics: Martensitic transformations in crack-tip stress fields, J. Am. Ceram. Soc., 63 (1980), pp. 241-248

Garcia et al., 2002: M.A. Garcia, S.E. Paje, J. Llopis, Relationship between mechanical grinding and photoluminescence of zirconia-toughened-alumina ceramics, Mater. Sci. Eng. A, 325 (2002), pp. 302-306

Garvie and Nicholson, 1972: R.C. Garvie, P.S. Nicholson, Phase analysis in zirconia systems, J. Am. Ceram. Soc., 55 (1972), pp. 303-305

Gutknecht et al., 2007: D. Gutknecht, J. Chevalier, V. Garnier, G. Fantozzi, Key role of processing to avoid low temperature ageing in alumina zirconia composites for orthopaedic application, J. Eur. Ceram. Soc., 27 (2007), pp. 1547-1552

He and Clarke, 1995: J. He, D.R. Clarke, Determination of the piezospectroscopic coefficients for chromium-doped sapphire, J. Am. Ceram. Soc., 78 (1995), pp. 1347-1353

Ma and Clarke, 1993: Q. Ma, D.R. Clarke, Stress measurement in single-crystal and polycrystalline ceramics using their optical fluorescence, J. Am. Ceram. Soc., 76 (1993), pp. 1433-1440

Ma and Clarke, 1994: Q. Ma, D.R. Clarke, Piezospectroscopic determination of residual stresses in polycrystalline alumina, J. Am. Ceram. Soc., 77 (1994), pp. 298-302

Merlani et al., 2001: E. Merlani, C. Schmid, V. Sergo, Residual stresses in alumina/zirconia composites: effect of cooling rate and grain size, J. Am. Ceram. Soc., 84 (2001), pp. 2962-2968

Ortiz and Suresh, 1993: M. Ortiz, S. Suresh, Statistical properties of residual stresses and intergranular fracture in ceramic materials, J. Appl. Mech., 60 (1993), pp. 77-84

Pecharroman et al., 2003: C. Pecharroman, J.F. Bartolome, J. Requena, J.S. Moya, S. Deville, J. Chevalier, G. Fantozzi, R. Torrecillas, Percolative mechanism of ageing in zirconia containing ceramics for biomedical applications, Adv. Mater., 15 (2003), pp. 507-511

Reichl and Steinbrench, 1988: A. Reichl, R. Steinbrench, Determination of crack-bridging forces in alumina, J. Am. Ceram. Soc., 71 (1988), pp. c299-c301

Rice and Pohanka, 1979: R.W. Rice, R.C. Pohanka, Grain size dependence of spontaneous cracking in ceramics, J. Am. Ceram. Soc., 62 (1979), pp. 559-563

Scattergood et al., 1991: R.O. Scattergood, S. Srinivasan, T.G. Bifano, R-curve effects for machining and wear of ceramics, Ceram. Acta., 3 (1991), pp. 53-64

Schehl et al., 2002 : M. Schehl, L.A. Diaz, R. Torrecillas, Alumina nanocomposites from powder-alkoxide mixtures, Acta Mater., 50 (2002), pp. 1125-1139

Selcuk and Atkinson, 2002: A. Selcuk, A. Atkinson, Analysis of the Cr3+ luminescence spectra from thermally grown oxide in thermal barrier coatings, Mater. Sci. Eng. A, 335 (2002), pp. 147-156

Sergo et al., 1998 : V. Sergo, G. Pezzotti, O. Sbaizero, T. Nishida, Grain size inflence on residual stresses in alumina/zirconia composites, Acta Mater., 46 (1998), pp. 1701-1710

Shyam and Lara-Curzio, 2006: A. Shyam, E. Lara-Curzio, The double-torsion testing technique for determination of fracture toughness and slow crack growth behavior of materials: a review, J. Mater. Sci., 41 (2006), pp. 4093-4104

Toraya et al., 1984 : H. Toraya, M. Yoshimura, S. Shigeyuki, Calibration curve for quantitative analysis of the monoclinic-tetragonal ZrO2 systems by X-ray diffraction, J. Am. Ceram. Soc., 67:C (1984), pp. 119-121

Wan et al., 1990: K.T. Wan, S. Lathabai, B.R. Lawn, Crack velocity functions and thresholds in brittle solids, J. Eur. Ceram. Soc., 6 (1990), pp. 259-268

Williams and Evans, 1973: D.P. Williams, A.G. Evans, A simple method to study slow crack growth, J. Test. Eval., 1 (1973), pp. 264-270

Willmann, 1998: G. Willmann, Ceramics for total hip replacement—what a surgeon should know, Orthopedics, 21 (2) (1998), pp. 173-177

Willmann, 2000: G. Willmann, Ceramic femoral head retrieval data, Clin. Orthop. Relat. Res., 379 (2000), pp. 22-28